\documentclass[preprint,showkeys]{revtex4} 
\usepackage{amsmath}
\usepackage{latexsym}
\usepackage{amssymb}
\usepackage{graphics,epstopdf}
\usepackage{amsmath,amssymb,amsthm}
\usepackage{graphicx}
\usepackage[colorlinks=true, citecolor=blue, urlcolor=blue ]{hyperref}
\usepackage{slashed}
\usepackage[title]{appendix}
\usepackage{cleveref}

\begin{document}
\title{Entanglement in phase-space distribution for an anisotropic harmonic oscillator in noncommutative space}

\author{Pinaki Patra}
\thanks{Corresponding author}
\email{monk.ju@gmail.com}
\affiliation{Department of Physics, Brahmananda Keshab Chandra College, Kolkata, India-700108}

\keywords{Anisotropic oscillator; Peres-Horodecki criterion; Wigner distribution; Noncommutative space; Coherent state; Symplectic diagonalization}

\date{\today}

\begin{abstract}
The bi-partite Gaussian state, corresponding to an anisotropic harmonic oscillator in a noncommutative space (NCS), is investigated with the help of Simon's separability condition (generalized Peres-Horodecki criterion). It turns out that, to exhibit the entanglement between the noncommutative coordinates, the parameters (mass and frequency) have to satisfy a unique constraint equation. \\
We have considered the most general form of an anisotropic oscillator in  NCS, with both spatial and momentum non-commutativity. The system is transformed to the usual commutative space (with usual Heisenberg algebra) by a well-known Bopp shift.  The system is transformed into an equivalent simple system by a unitary transformation, keeping the intrinsic symplectic structure ($Sp(4,\mathbb{R})$) intact. Wigner quasiprobability distribution is constructed for the bipartite Gaussian state with the help of a Fourier transformation of the characteristic function. It is shown that the identification of the entangled degrees of freedom is possible by studying the Wigner quasiprobability distribution in phase space. We have shown that the coordinates are entangled only with the conjugate momentum corresponding to other coordinates. 
  
\end{abstract}
 
 \maketitle
\section{Introduction}
There is a consensus that most of the theories of quantum gravity appear to predict departures from classical gravity only at energy scales on the order of $10^{19}$ GeV \cite{qg1,qg2} (By way of comparison, the LHC was designed to run at a maximum collision energy of $14$  TeV \cite{lhc}). At that energy scale, the space-time structure is believed to be deformed in such a manner that the usual notion of commutative space will be ceased \cite{csseased1,csseased2}.
It is almost a Gospel that the fundamental concept of space-time is mostly compatible with quantum theory in noncommutative space (NCS) \cite{Witten,Juan}. Formalisms of Quantum theory in NCS have been successfully applied to various physical situations \cite{ncsho,ncsosc,ncstopics,ncsfluid}. For the reviews on NCS, one can see \cite{ncsreview1,ncsreview2}.  \\
Perhaps, due to the present-day technological limitation of attainable energy, the unified quantum theory of gravity lacks any direct experimental evidence. Moreover, the huge gap between the required ($10^{19}$ GeV) and the currently achievable energy scales suggests that we have to wait a long to successfully perform quantum gravity (QG) experiments in the regime of present-day colliders.
However, instead of building a larger collider, there are various suggestions for the passive experiments regarding the QG, with the help of interferometry
\cite{gravitationalwave1,catstatedeformed1,catstatedeformed2,catstatedeformed3},  in particular with the utilization of Coherent states (CS), for which the maximum contrast in the interference pattern is achieved \cite{interfero1}.  In the context of CS, the bipartite Gaussian states are particularly important for their frequent appearance in systems with small oscillations \cite{bipartitegaussian1,bipartitegaussian2}. For example, the vibration of the test mass (corresponding to the masses on which the interferometric mirrors are placed) under the effects of the tidal forces of gravitational waves, can be demonstrated by an anisotropic oscillator \cite{gravitationalwave1,gravitationalwave2,gravitationalwave3,gravitationalwave4}, which exhibits a bipartite Gaussian state, upon consideration in a two-dimensional noncommutative space \cite{entangledgaussian1,entangledgaussian2,entangledgaussian3,entangledgaussian4}. \\ 
The entanglement properties of the bipartite state, corresponding to the noncommutative spatial degrees of freedom, are the central theme of the present article. A general separability criterion for a bipartite Gaussian state was proposed in \cite{separability1}, which is a generalization for the continuous variable states, over the discrete qubits \cite{separability2,separability3}. \\
At first, using a Bopp’s shift in co-ordinates \cite{boppsshift1,boppsshift2,boppsshift3,boppsshift4}, we have transformed the Hamiltonian ($\hat{H}_{nc}$) of the anisotropic harmonic oscillator in noncommutative space, to the equivalent Hamiltonian ($\hat{H}_c$) in commutative space. The structure of the $\hat{H}_c$ is similar to the Hamiltonians corresponding to a charged particle moving in presence of an inhomogeneous magnetic field. One of the major computational difficulties in the construction of the CS structure of an anisotropic oscillator in a magnetic field lies in the diagonalization of the quadratic system, keeping the intrinsic symplectic structure intact. In this article, to construct a  linear canonical transformation (the group $Sp(4, \mathbb{R})$) to reduce our Hamiltonian to a simplified representative one, we have revisited the formalisms stated in \cite{separability4,separability5,separability6,separability7,separability8}.  The ground state of the anisotropic harmonic oscillator (AHO) turns out to be a bipartite Gaussian state. Utilizing Simon's separability criterion \cite{separability1}, we have obtained the restrictions on the masses ($m_1,m_2$) and the parameters ($\theta,\eta$) (minimal length) of the noncommutative coordinates and momentums, for which the CS is separable. \\
For an entangled state, the identification of the entangled degrees of freedom is not straightforward. With the help of the well-known Wigner quasiprobability distribution (WQD)  \cite{wigner1,wigner2,wigner3,wigner4,wigner5,wigner6}, we have identified the coordinates, which are entangled. Since WQD is the Fourier transform of the expectation value of the characteristic function of the random variables, it offers a phase-space description of a quantum system. It is believed that one can conceptually identify where the quantum corrections enter a problem by comparing it with the classical version. However, philosophical debates in this regard are inevitable, which we shall avoid in our present article. Rather, we shall indulge ourselves in the computation of WQD for our problem, keeping in mind that, being entirely real, WQD simplifies both the calculation and the interpretation, of the results. In our case, the random variables are the co-ordinate $(\hat{x}_1,\hat{x}_2)$ and momentum $(\hat{p}_1,\hat{p}_2)$ operators in commutative space. The study of the WQD suggests that for the system under consideration (AHO in NCS) coordinate degrees of freedom are entangled with the conjugate momentum corresponding to another coordinate. In particular, $\hat{x}_1$ is entangled with $\hat{p}_2$ and $\hat{x}_2$ is entangled with $\hat{p}_1$.\\
The organization of the article is the following. First, we have briefly discussed the system under consideration. Then, we diagonalized the Hamiltonian, with the help of a $Sp(4,\mathbb{R})$ canonical transformation. Coherent states are formed from the annihilation operators. After that, Simon's separability criterion is utilized to identify the constraints on the parameters, which can be used as an entanglement criterion for our system. The identification of the entangled degrees of freedom was performed with the help of the study of Wigner quasiprobability phase space distribution.
\section{Anisotropic harmonic oscillator in noncommutative space}
We have studied the phenomenology of a two-dimensional anisotropic harmonic oscillator in noncommutative (NC) space, with both the configuration and momentum space non-commutativity.  We shall consider the following Bopp’s shift \cite{boppsshift1,boppsshift2,boppsshift3,boppsshift4}, which connects the NC-space co-ordinates ($\hat{\tilde{X}}_1,\hat{\tilde{X}}_2$) and momentum ($\hat{\tilde{P}}_1,\hat{\tilde{P}}_2$) to the commutative space co-ordinates ($\hat{x}_1,\hat{x}_2$) and momentum ($\hat{p}_1,\hat{p}_2$).
\begin{eqnarray}\label{boppshift}
\begin{array}{cc}
  \hat{\tilde{X}}_1 = l_0(\hat{x}_1-\frac{\theta}{2\hbar}\hat{p}_2 ), &
\hat{ \tilde{P}}_1 = l_0(\hat{p}_1 + \frac{\eta}{2\hbar}\hat{x}_2), \\
\hat{\tilde{X}}_2 =l_0( \hat{x}_2+ \frac{\theta}{2\hbar}\hat{p}_1) , &
 \hat{\tilde{P}}_2 =l_0(\hat{ p}_2 - \frac{\eta}{2\hbar}\hat{x}_1 ).
\end{array}
\end{eqnarray}
In \cite{boppsshift3}, it was shown that the noncommutative algebra is invariant under scale factor $l_0$, so that all values of  $l_0$ describe the same physical model. For simplicity, we shall consider $l_0=1$ in our present study.
With this choice, the NC extension of quantum mechanics in two dimensions is realized by the following commutation relations.
\begin{eqnarray}
\begin{array}{ccc}
 [\hat{\tilde{X}}_1,\hat{\tilde{X}}_2]= i\theta, & [\hat{\tilde{P}}_1,\hat{\tilde{P}}_2] = i\eta, & [\hat{\tilde{X}}_i,\hat{\tilde{P}}_j] = i\hbar_{e}\delta_{ij},
\end{array} \label{commutationrelation}\\
\mbox{with}\; \hbar_{e}=(1+\frac{\theta\eta}{4\hbar^2})\hbar .
\end{eqnarray}
Whereas, the usual Heisenberg algebra for the commutative space reads
\begin{eqnarray}
 [\hat{x}_1,\hat{x}_2]=[\hat{p}_1,\hat{p}_2]=0,\; [\hat{x}_i, \hat{p}_j] = i\hbar \delta_{ij}.
\end{eqnarray}
Where the Kronecker delta ($\delta_{ij}$) is defined as
\begin{eqnarray}
 \delta_{ij}=\left\{\begin{array}{cc}
                     1, & i=j \\
                     0, & i\neq j
                    \end{array}
\right. \;\;\; ;i,j=1,2.
\end{eqnarray}
We would like to mention that the noncommutative parameters $\theta$, $\eta$, and the effective Planck constant $\hbar_e = (1+\frac{\theta\eta}{4\hbar^2})\hbar$ are experimentally measurable quantities for the scale factor $l_0=1$  \cite{boppsshift3,boppsshift4}.\\
In this NC space (NCS), let us consider, a two dimensional anisotropic harmonic oscillator (AHO), which is described by the Hamiltonian 
\begin{equation}\label{hamiltoniannc}
 \hat{H}_{nc}=\frac{1}{2m_1}\hat{\tilde{P}}_1^2 + \frac{1}{2m_2}\hat{\tilde{P}}_2^2 + \frac{1}{2}m_1 \tilde{\omega}_1^2 \hat{\tilde{X}}_1^2 + \frac{1}{2}m_2 \tilde{\omega}_2^2\hat{\tilde{X}}_2^2 .
\end{equation}
Without loss of generality, we shall consider the mass ($m_1,m_2$) and the frequencies ( $\tilde{\omega}_1, \tilde{\omega}_2$) are constants and can have only positive values. We would like to mention that the Hamiltonian ~\eqref{hamiltoniannc}  has different masses in different directions as well as different
harmonic potentials. Thus, it is anisotropic in two ways.\\
Although the linear transformation~\eqref{boppshift}  changes the symplectic structure (thus, not unitary), it enables us to convert the Hamiltonian in the NCS into a modified Hamiltonian in the commutative
space having an explicit dependence on the deformation parameters $\theta,\eta$. Then the
states of the system are functions on the ordinary Hilbert space.
The dynamics of the system are now governed by the Schr\"{o}dinger equation
with the NC parameters $\theta$ and $\eta$ dependent Hamiltonian, which reads
\begin{equation}\label{Hamiltonian}
 \hat{H}= \frac{1}{2\mu_1}\hat{p}_1^2 + \frac{1}{2\mu_2}\hat{p}_2^2 + \frac{1}{2}\mu_1\omega_1^2 \hat{x}_1^2 + \frac{1}{2}\mu_2\omega_2^2 \hat{x}_2^2 + \hat{L}_d ,
\end{equation}
with
\begin{equation}
 \hat{L}_d =  \nu_1 \left\{\hat{x}_2,\hat{p}_1\right\} - \nu_2 \left\{ \hat{x}_1,\hat{p}_2  \right\}.
\end{equation}
New parameters ($\mu_1,\mu_2,\omega_1,\omega_2,\nu_1,\nu_2$ are connected with the previous ($m_1,m_2,\tilde{\omega}_1,\tilde{\omega}_2$) one by
\begin{eqnarray}
 \frac{1}{\mu_1} &=& \frac{1}{m_1}+ \frac{1}{4\hbar^2} m_2\tilde{\omega}_2^2\theta^2, \\
 \frac{1}{\mu_2} &=& \frac{1}{m_2}+ \frac{1}{4\hbar^2} m_1\tilde{\omega}_1^2\theta^2,\\
 \mu_1\omega_1^2 &=& m_1\tilde{\omega}_1^2 + \frac{\eta^2}{4\hbar^2 m_2} ,\\
 \mu_2\omega_2^2 &=& m_2\tilde{\omega}_2^2 + \frac{\eta^2}{4\hbar^2 m_1},\\
 \nu_1 & =& \frac{1}{4m_1\hbar} (\eta + m_1 m_2\tilde{\omega}_2^2 \theta),\\
 \nu_2 &=& \frac{1}{4m_2\hbar} (\eta + m_1 m_2\tilde{\omega}_1^2 \theta).
\end{eqnarray}
It is easy to see that the deformed angular momentum operator $\hat{L}_d$ will be reduced to the usual angular momentum operator for an isotropic oscillator (IHO). \\
 On the way of comparison of our system with an anisotropic oscillator in a magnetic field, we note that an anisotropic  oscillator (potential $V_{an}(x_1,x_2)=k_1x_1^2/2 + k_2x_2^2/2$) with charge $e$ and effective mass  $M=(\mu_1,\mu_2)$, under the magnetic vector potential 
 \begin{equation}\label{vecpotential}
  \vec{\mathcal{A}}=(-\alpha_{01}x_2,\alpha_{02} x_1),
 \end{equation}
 corresponds to the Hamiltonian
\begin{equation}
 \hat{H}_{mag}= \frac{1}{2\mu_1}\hat{p}_1^2 + \frac{1}{2\mu_2}\hat{p}_2^2 + \frac{1}{2}\alpha_1 \hat{x}_1^2 + \frac{1}{2}\alpha_2 \hat{x}_2^2 + \hat{L}_d .
\end{equation}
Since in $2$-dimension curl produces a pseudo-scalar, the vector potential ~\eqref{vecpotential} corresponds to the homogeneous magnetic field
\begin{equation}
 B_{h}= \alpha_{01}+\alpha_{02}.
\end{equation}
One can identify that our system is equivalent to the anisotropic charged oscillator in the magnetic field $B_h$ with the parameter values
\begin{eqnarray}
\begin{array}{ccc}
 \alpha_1= k_1+ \frac{e^2\alpha_{02}^2}{\mu_2 c^2}, & \alpha_2= k_2+ \frac{e^2\alpha_{01}^2}{\mu_1 c^2}, & \\
 \nu_1=\frac{e\alpha_{01}}{2\mu_1 c}, &\nu_2=\frac{e\alpha_{02}}{2\mu_2 c}, & B_h= \frac{2c}{e}(\mu_1\nu_1+\mu_2\nu_2),
\end{array}
\end{eqnarray}
$c$ being the speed of light in free space.\\
Moreover, one can see that these types of systems are also equivalent to the two-dimensional Maxwell-Chern-Simon's model in long wavelength limit, which has a potential application in the theory of anyons \cite{Dunne}.\\
For an IHO,  the usual angular momentum operator ($\hat{L}$) commutes with the Hamiltonian.  Thus the problem can be reduced to the two noninteracting harmonic oscillators. However, for AHO ($\nu_1\neq \nu_2$), the process of diagonalization is not straightforward. In the next section, the diagonalization of~\eqref{Hamiltonian} is illustrated.
\section{diagonalization of the system}
The Hamiltonian~\eqref{Hamiltonian} can be expressed in the following quadratic form.
\begin{equation}
 \hat{H}=\frac{1}{2}X^T \hat{\mathcal{H}}X,
\end{equation}
where 
\begin{eqnarray}\label{Xdefn}
 X= \left(\hat{X}_1,\hat{X}_2,\hat{X}_3,  \hat{X}_4\right) =  \left(\hat{x}_1,
  \hat{p}_1,
   \hat{x}_2,
    \hat{p}_2
 \right)^T, \;  \hat{\mathcal{H}} = \left( \begin{array}{cc}
 \hat{C} & \hat{A}^T \\
 \hat{A} & \hat{B}
\end{array}\right),\\
\mbox{with}\; \;
\hat{B} = \left( \begin{array}{cc}
 \mu_2\omega_2^2 & 0 \\
 0 & \frac{1}{\mu_2}
\end{array}\right),\;
\hat{C} = \left( \begin{array}{cc}
 \mu_1\omega_1^2 & 0 \\
 0 & \frac{1}{\mu_1}
\end{array}\right),\;
\hat{A}= \left( \begin{array}{cc}
 0 & 2\nu_1 \\
 -2\nu_2 & 0
\end{array}\right).
\end{eqnarray}
 $X^T$ denotes the matrix transposition of $X$.
One can identify that
\begin{eqnarray}\label{symplecticdefn}
 \hat{\mathcal{S}}\hat{\mathcal{J}}_4 + \hat{\mathcal{J}}_4 \hat{\mathcal{S}}^T =0,
\end{eqnarray}
where
\begin{eqnarray}\label{symplecticmatrix}
 \hat{\mathcal{S}}= \hat{\mathcal{J}}_4\hat{\mathcal{H}},\;\;
\mbox{with}\;\;
 \hat{\mathcal{J}}_{2n}= \left( \begin{array}{cc}
0 & \hat{\mathbb{I}}_n \\
-\hat{\mathbb{I}}_n & 0
\end{array}\right),
\end{eqnarray}
$\hat{\mathbb{I}}_n$ being the $n\times n$ identity matrix.
From the equation ~\eqref{symplecticdefn}, we can conclude that $\mathcal{S}$ is a symplectic matrix ($\mathcal{S} \in Sp (4,\mathbb{R})$).
Since our co-ordinates are the Cartesian type, the natural range
(spectrum) for each of them is the entire real line.   Moreover, the commutation relations 
\begin{equation}\label{columncommutation}
 [\hat{X}_\alpha, \hat{X}_\beta] =-( \Sigma_y)_{\alpha\beta};\; (\alpha,\beta=1,2,3,4),
\end{equation}
generate an intrinsic symplectic matrix
\begin{eqnarray}\label{Sigmay}
 \hat{\Sigma}_y=\mbox{diag}(\sigma_y,\sigma_y),
\end{eqnarray}
where the Pauli matrices are given by
\begin{eqnarray}
 \sigma_x= \left(\begin{array}{cc}
                                0 & 1\\
                                1 & 0
                               \end{array}
\right), \; 
\sigma_y =\left(\begin{array}{cc}
                  0 & -i \\
                  i & 0
                 \end{array}
\right), \; 
\sigma_z= \left(\begin{array}{cc}
                                1 &0\\
                                0&-1
                               \end{array}
\right).
\end{eqnarray}
Here we have considered $\hbar=1$, which will be followed throughout our present discussion.\\
The present section aims to diagonalize $\hat{H}$ keeping the symplectic structure ~\eqref{columncommutation} intact.\\
The structure constants of the closed quasi-algebra of the co-ordinates ($\hat{X}_i,i=1,..,4.$) with respect to the Hamiltonian ($\hat{H}$), induce the following commutation relation.
\begin{equation}
 \left[i \hat{H}, X\right] =\hat{\Omega} X,
\end{equation}
where
\begin{equation}\label{OmegaHconnection}
 \hat{\Omega}= i \hat{\Sigma}_y \hat{\mathcal{H}}.
\end{equation}
The next step is to diagonalize the matrix  $\hat{\Omega}$. We would like to relegate the detailed calculation of the diagonalization of $\hat{\Omega}$ to Appendix A.\\
The diagonal Hamiltonian is given by
\begin{equation}\label{HDfactorization}
 \hat{H}_D= \frac{1}{2}\zeta^\dagger \hat{\Sigma}\zeta=\lambda_1 \left(\hat{\zeta}_1^\dagger \hat{\zeta}_1 +\frac{1}{2} \right) + \lambda_2 \left(\hat{\zeta}_2^\dagger \hat{\zeta}_2 +\frac{1}{2} \right),
 \end{equation}
 where $\hat{\Sigma}
  = \mbox{diag}(\lambda_1,\lambda_1,\lambda_2,\lambda_2)$.
The explicit forms of $\zeta_i,\; i=1,2$ and $\zeta_i^\dagger,\; i=1,2$, along with their algebra are given in subsections C and D of Appendix-A. The detailed calculations of the four distinct purely imaginary eigenvalues ($\mp i\lambda_1,\mp i\lambda_2$) of $\hat{\Omega}$ are given in the subsection-B  of the Appendix-A. \\
Next section deals with the separability of the bipartite state of our system.
\section{Separability of the bipartite state}
The factorized form ~\eqref{HDfactorization} of $\hat{H}_D$ suggests that the states of the system may be expressed as
\begin{eqnarray}\label{statesn1n2}
 \vert n_1,n_2\rangle = \frac{1}{\sqrt{n_1!n_2!}} (\hat{\zeta}_1^\dagger)^{n_1} (\hat{\zeta}_2^\dagger)^{n_2}\vert 0,0\rangle ; n_1,n_2=0,1,2,3,.....
\end{eqnarray}
The corresponding energy eigenvalues are
\begin{equation}\label{energyn1n2}
 E_{n_1,n_2}=\left(n_1+\frac{1}{2}\right)\lambda_1 + \left(n_2+\frac{1}{2}\right)\lambda_2.
\end{equation}
For our purpose, we shall only need the ground state, which is given by
\begin{equation}
 \hat{\zeta}_1\vert 0,0\rangle =\hat{\zeta}_2\vert 0,0\rangle =0.
\end{equation}
A $2$-mode pure Gaussian state ($ \vert \alpha,z\rangle$) is obtained
from the $2$-mode vacuum state $\vert 0,0\rangle$ by \cite{gaussian1,gaussian2}
\begin{equation}
 \vert \alpha,z\rangle =\hat{S}_2(z)\hat{D}_2(\alpha)\vert 0,0\rangle,
\end{equation}
where the displacement operator $\hat{D}_2(\alpha)=e^{\alpha^T \zeta_{a_{\dagger}}-\alpha^\dagger\zeta_{a}}$ and the squeezed operator $\hat{S}_2(z)=e^{\frac{1}{2}(\zeta_a^\dagger z  \zeta_{a_{\dagger}}-\zeta_a^T z^\dagger \zeta_a)}$ are given through $\zeta_a=(\hat{\zeta}_1,\hat{\zeta}_2)^T$, $\zeta_{a_{\dagger}}=(\hat{\zeta}_1^\dagger,\hat{\zeta}_2^\dagger)^T$, along with the displacement vector $\alpha$ and a complex symmetric squeeze matrix $z$. It is evident that $\vert 0,0\rangle$ is a two mode Gaussian state with $\alpha^T=(0,0),z=\hat{0}_{2\times 2}$. The advantage to use Gaussian states is that they are entirely characterized by the covariance matrix. It means that typical issues of continuous variables quantum information theory, which are generally difficult to handle in an
infinite Hilbert space, can be faced up with the help of finite matrix theory \cite{gaussian3}. \\
The explicit form of $\vert 0,0\rangle $ in position $\{\vert x\rangle\}$ representation  reads
\begin{equation}\label{psientangle}
 \psi_0(x_1,x_2)=\langle x_1,x_2\vert 0,0\rangle = A_0 e^{-\Lambda_{11}x_1^2-\Lambda_{22}x_2^2-\Lambda_{12}x_1x_2}.
\end{equation}
Where
\begin{eqnarray}
 \Lambda_{11} &=& \frac{\mu_1 \mu_2  \tau_\lambda \sqrt{\Delta_\lambda}}{4\delta_\lambda}, \label{a}\\
 \Lambda_{22} &=& \frac{\mu_2 \tau_\lambda ( \mu_2\omega_2^2 - 4\mu_1\nu_1^2)}{4\delta_\lambda}, \label{b}\\
\Lambda_{12} &=& \frac{2i\mu_2}{\delta_\lambda} (4\mu_1 \nu_1^2\nu_2 - \mu_2\nu_2 \omega_2^2 +\mu_1\nu_1\sqrt{\Delta_\lambda}), \label{c}
\end{eqnarray}
with
\begin{equation}
 \delta_\lambda = \mu_2(\omega_2^2 + \sqrt{\Delta_\lambda})-4\mu_1\nu_1^2.
\end{equation}
$\tau_\lambda$ and $\Delta_\lambda$ are the trace and determinant of $\hat{\Sigma}$, respectively. 
The normalization constant is given by $\vert A_0\vert =\sqrt[4]{4 \Lambda_{11}\Lambda_{22}/\pi^2}$.\\
It is worth noting that $\Lambda_{11}$ and $\Lambda_{22}$ are real, whereas $\Lambda_{12}$ is purely imaginary, which makes the computation of expectation values very simple. In particular,
\begin{eqnarray}\label{expectationvalues}
 \begin{array}{cccc}
  \langle \hat{x}_1\rangle =0, & \langle \hat{x}_1^2\rangle =\frac{1}{4\Lambda_{11}}, & \langle \hat{x}_2\rangle =0, & \langle \hat{x}_2^2\rangle =\frac{1}{4\Lambda_{22}}, \\
  \langle \hat{p}_1\rangle =0, & \langle \hat{p}_1^2\rangle =\frac{d}{4\Lambda_{22}}, & \langle \hat{p}_2\rangle =0, & \langle \hat{p}_2^2\rangle =\frac{d}{4\Lambda_{11}},\\
  \langle \hat{x}_1 \hat{x}_2\rangle =0, & \langle \hat{p}_1 \hat{p}_2\rangle =0, & \langle \hat{x}_1 \hat{p}_2 \rangle =\frac{i\Lambda_{12}}{4\Lambda_{11}}, & \\
  \langle \left\{ \hat{x}_1, \hat{p}_1 \right\} \rangle =0, & \langle \left\{ \hat{x}_2, \hat{p}_2 \right\} \rangle =0,& \langle \hat{p}_1 \hat{x}_2 \rangle =\frac{i\Lambda_{12}}{4\Lambda_{22}}. & 
 \end{array}\;; d= 4\Lambda_{11}\Lambda_{22}-\Lambda_{12}^2.
\end{eqnarray}
Since $\Lambda_{12}$ is purely imaginary, all the expectation values in ~\eqref{expectationvalues} are real. Moreover, the variances of the observables give
\begin{equation}
 \Delta \hat{x}_1\Delta \hat{p}_1=\Delta \hat{x}_2\Delta \hat{p}_2= \frac{1}{2}\sqrt{1+ \frac{\vert \Lambda_{12}\vert^2}{4\Lambda_{11}\Lambda_{22}}},
\end{equation}
which means the state $\psi_0$ is a squeezed state for nonzero deformation parameters. However, for the usual commutative space ($\Lambda_{12}=0$) the squeezing goes off and $\psi_0$ is reduced to a coherent state.\\
To study the separability criterion, let us consider the variance matrix
\begin{eqnarray}
 \hat{V}= \left(\begin{array}{cc}
                  V_{11} & V_{12} \\
                 V_{12}^T & V_{22}
                \end{array}
\right),
\end{eqnarray}
with
\begin{eqnarray}
 V_{11}&=&\left(\begin{array}{cc}
                  \langle \hat{x}_1^2 \rangle & \langle \{ \hat{x}_1,\hat{p}_1 \} \rangle \\
                 \langle \{ \hat{x}_1,\hat{p}_1 \} \rangle  & \langle \hat{p}_1^2 \rangle
                \end{array}
\right), \\
 V_{22}&=& \left(\begin{array}{cc}
                  \langle \hat{x}_2^2 \rangle & \langle \{ \hat{x}_2,\hat{p}_2 \} \rangle \\
                 \langle \{ \hat{x}_2,\hat{p}_2 \} \rangle  & \langle \hat{p}_2^2 \rangle
                \end{array}
\right),\\
V_{12}&=& \left(\begin{array}{cc}
                  \langle \hat{x}_1\hat{x}_2 \rangle & \langle \hat{x}_1\hat{p}_2  \rangle \\
                 \langle \hat{p}_1\hat{x}_2  \rangle  & \langle \hat{p}_1\hat{p}_2  \rangle
                \end{array}
\right).
\end{eqnarray}
The necessary and sufficient condition for the separability of a bipartite Gaussian state reads \cite{separability1}
\begin{eqnarray}\label{simonscriterion}
 \Delta_1\Delta_2 + \left(\frac{1}{4}- \vert \Delta_{12}\vert\right)^2 -\mbox{Trace}(V_{11}\tilde{V}_{12}V_{22}\tilde{V}_{21}) \ge (\Delta_1+ \Delta_2).
\end{eqnarray}
Where $\Delta_1,\; \Delta_2,\; \Delta_{12}$ are the determinant of $V_{11},\;V_{22}, V_{12}$ respectively, and
\begin{eqnarray}
 \tilde{V}_{ij}= J_2 V_{ij}J_2,\;\;
 V_{21}=V_{12}^T.
\end{eqnarray}
The symplectic matrix $J_2$ is given by ~\eqref{symplecticmatrix}. \\
Using the explicit forms ~\eqref{expectationvalues} in ~\eqref{simonscriterion}, we can see that the bipartite state is separable only for 
\begin{equation}\label{constraintseparability}
 \Lambda_{12}^2 \ge 0.
\end{equation}
However, $\Lambda_{12}$ is purely imaginary. Therefore,  ~\eqref{constraintseparability} holds only for the  case 
\begin{equation}\label{Lambda120}
 \Lambda_{12}=0.
\end{equation}
~\eqref{Lambda120} is satisfied with one of the following conditions.
\begin{itemize}
\item $\theta=\eta=0$: Corresponding to a 2-D anisotropic harmonic oscillator in commutative-space. 
 \item $\tilde{\omega}_1 =\tilde{\omega}_2$: The anisotropy due to the mass and the noncommutative parameters.
 \item  If  $m_1,m_2$ and $\tilde{\omega}_1, \tilde{\omega}_2$ satisfy 
 \begin{equation}\label{thetaetaconstraint}
        \theta m_1\tilde{\omega}_1 =\frac{\eta}{m_2 \tilde{\omega}_2}.
\end{equation}

\end{itemize}
In other words, for nonzero values of the noncommutative parameters (nonzero $\theta,\eta$) the bipartite states are entangled except for the parameter values ~\eqref{thetaetaconstraint}. In particular, this entanglement behavior solely depends on the noncommutative parameters, which is in agreement with the result of \cite{entangledgaussian1}. However, we would like to mention that, in \cite{entangledgaussian1}, only the position-position noncommutativity was considered, whereas, we have considered both position-position and momentum-momentum non-commutativity.  That means, our result is a generalization of the previously reported results. In the following subsection, we have shown that this type of entanglement is generated due to the noncommutative geometry.
\subsection{The entanglement is induced due to the noncommutative geometry}
If we consider the limit $\theta\to 0, \eta\to 0$, the eigenvalues of $\hat{\Omega}$ is reduecd to
\begin{equation}
 \lambda= \{\pm i\omega_1,\pm i\omega_2 \},
\end{equation}
which means
\begin{eqnarray}
 \begin{array}{ccc}
  \tau_\lambda = 2(\omega_1+\omega_2), & \Delta_\lambda = \omega_1^2 \omega_2^2,  & \delta_\lambda = \mu_2\omega_2(\omega_1+\omega_2), \\
  \Lambda_{11}= \frac{1}{2}\mu_1\omega_1, & \Lambda_{22}= \frac{1}{2}\mu_2\omega_2, & \Lambda_{12}= 0.
 \end{array}
\end{eqnarray}
The expectation values then reduced to
\begin{eqnarray}\label{expectationvaluesspecial}
 \begin{array}{cccc}
  \langle \hat{x}_1\rangle =0, & \langle \hat{x}_1^2\rangle =\frac{1}{2\mu_1\omega_1}, & \langle \hat{x}_2\rangle =0, & \langle \hat{x}_2^2\rangle =\frac{1}{2\mu_2\omega_2}, \\
  \langle \hat{p}_1\rangle =0, & \langle \hat{p}_1^2\rangle =\frac{1}{2}\mu_1\omega_1, & \langle \hat{p}_2\rangle =0, & \langle \hat{p}_2^2\rangle =\frac{1}{2}\mu_2\omega_2,\\
  \langle \hat{x}_1 \hat{x}_2\rangle =0, & \langle \hat{p}_1 \hat{p}_2\rangle =0, & \langle \hat{x}_1 \hat{p}_2 \rangle =0, & \\
  \langle \left\{ \hat{x}_1, \hat{p}_1 \right\} \rangle =0, & \langle \left\{ \hat{x}_2, \hat{p}_2 \right\} \rangle =0,& \langle \hat{p}_1 \hat{x}_2 \rangle =0. & 
 \end{array}
\end{eqnarray}
That means,  $\hat{x}_1$ and $\hat{p}_2$, as well as $\hat{x}_2$ and $\hat{p}_1$ are not correlated for usual commutative space. However,  from ~\eqref{expectationvalues}, we can see that the expectation values of $\hat{x}_1\hat{p}_2$ and $\hat{p}_1\hat{x}_2$ are nonzero for nonzero parameters of noncommutative geometry ($\theta,\eta$). This indicates the correlations between $\hat{x}_1$ and $\hat{p}_2$, as well as $\hat{x}_2$ and $\hat{p}_1$.  This type of entanglement is due to the noncommutative geometry.\\ 
In the next section, we have illustrated this fact with the help of the Wigner quasiprobability distribution.
\section{Wigner Distribution}
  The characteristic function ($\hat{M}(\vec{\tau},\vec{\theta})$) of the random variable  $X=(\hat{x}_1,\hat{p}_1,\hat{x}_2,\hat{p}_2),$ is defined by
\begin{equation}
 \hat{M}(\vec{\tau},\vec{\theta})= e^{i\mbox{Trace}(t^T X)}.
\end{equation}
with the parameter
\begin{equation}
 t=(\tau_1,\theta_1,\tau_2,\theta_2).
\end{equation}
The Fourier transformation of the expectation value of $\hat{M}$ is known as the Wigner quasiprobability distribution (WQD). For example, the WQD for a quantum state $\psi(x_1,x_2)$ may be given by
\begin{eqnarray}\label{wignerdefn}
 W(x_1,p_1,x_2,p_2)= \frac{2}{\pi\hbar^2} \int_{-\infty}^{\infty} \int_{-\infty}^{\infty} [\psi^*(x_1-t_1,x_2-t_2) e^{-\frac{2i}{\hbar}(t_1p_1+t_2p_2)} \nonumber \\ \psi(x_1+t_1,x_2+t_2) ]dt_1dt_2 .
\end{eqnarray}
Using~\eqref{psientangle} in ~\eqref{wignerdefn} and considering ~\eqref{a},~\eqref{b},~\eqref{c} and ~\eqref{expectationvalues}, the direct computation produces  the WQD for the anisotropic oscillator in NCS. In particular,
\begin{eqnarray}
 W(x_1,p_1,x_2,p_2)= \frac{2}{\pi\hbar^2} \exp(-2\langle \hat{p}_1^2 \rangle x_1^2 - 2\langle \hat{p}_2^2 \rangle x_2^2 -2\langle \hat{x}_1^2 \rangle p_1^2 -2\langle \hat{x}_2^2 \rangle p_2^2 \nonumber \\
 + 4 \langle \hat{x}_1\hat{p}_2 \rangle x_2p_1 + 4 \langle \hat{p}_1\hat{x}_2 \rangle x_1p_2).
\end{eqnarray}

\begin{figure}[!h]
\begin{minipage}[t]{0.5\textwidth}
\centering\includegraphics[width=\textwidth]{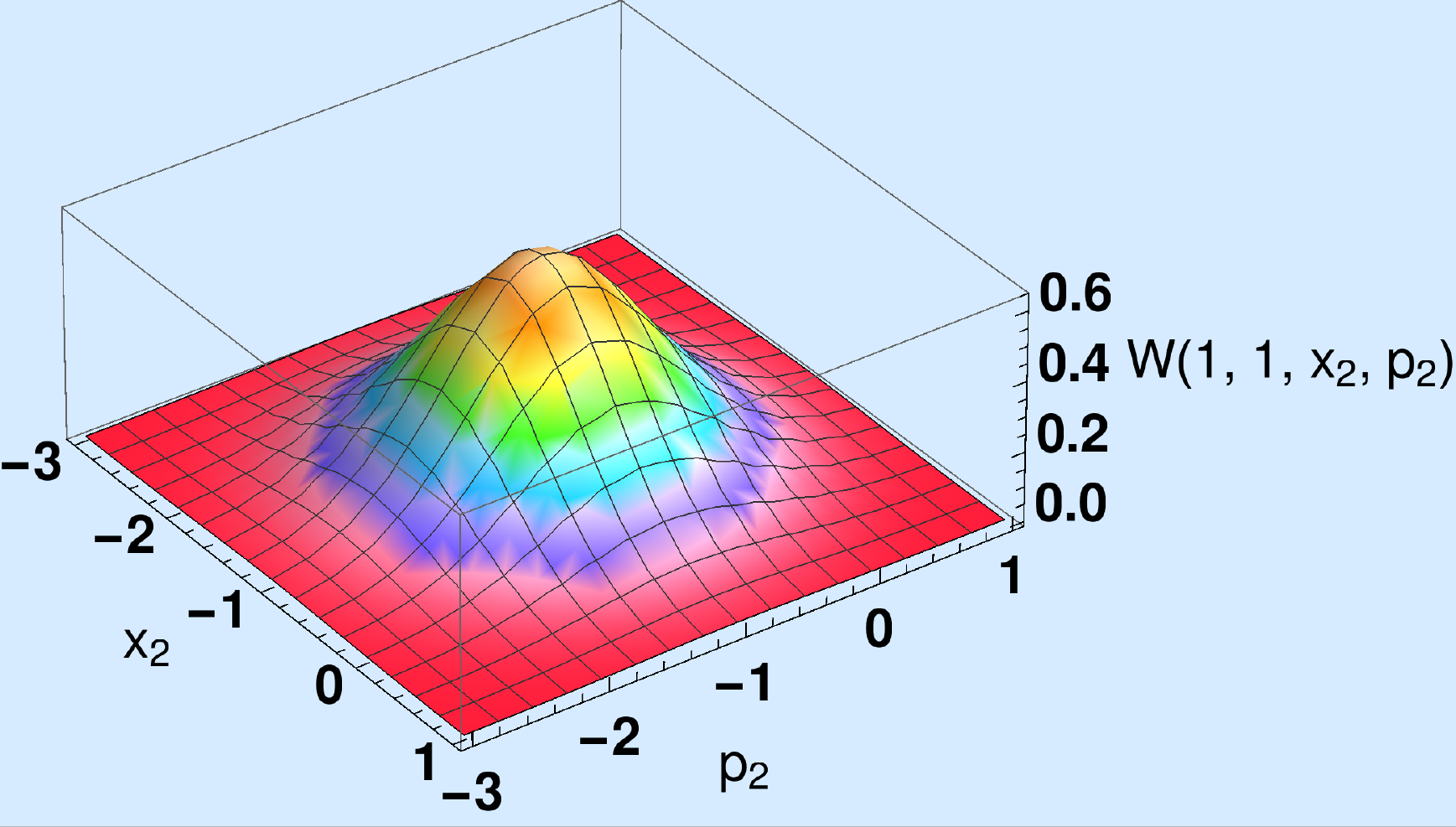}\\
{\footnotesize projection on $(x_1,p_1)=(1,1)$ plane.}
\end{minipage}\hfill
\begin{minipage}[t]{0.5\textwidth}
\centering\includegraphics[width=\textwidth]{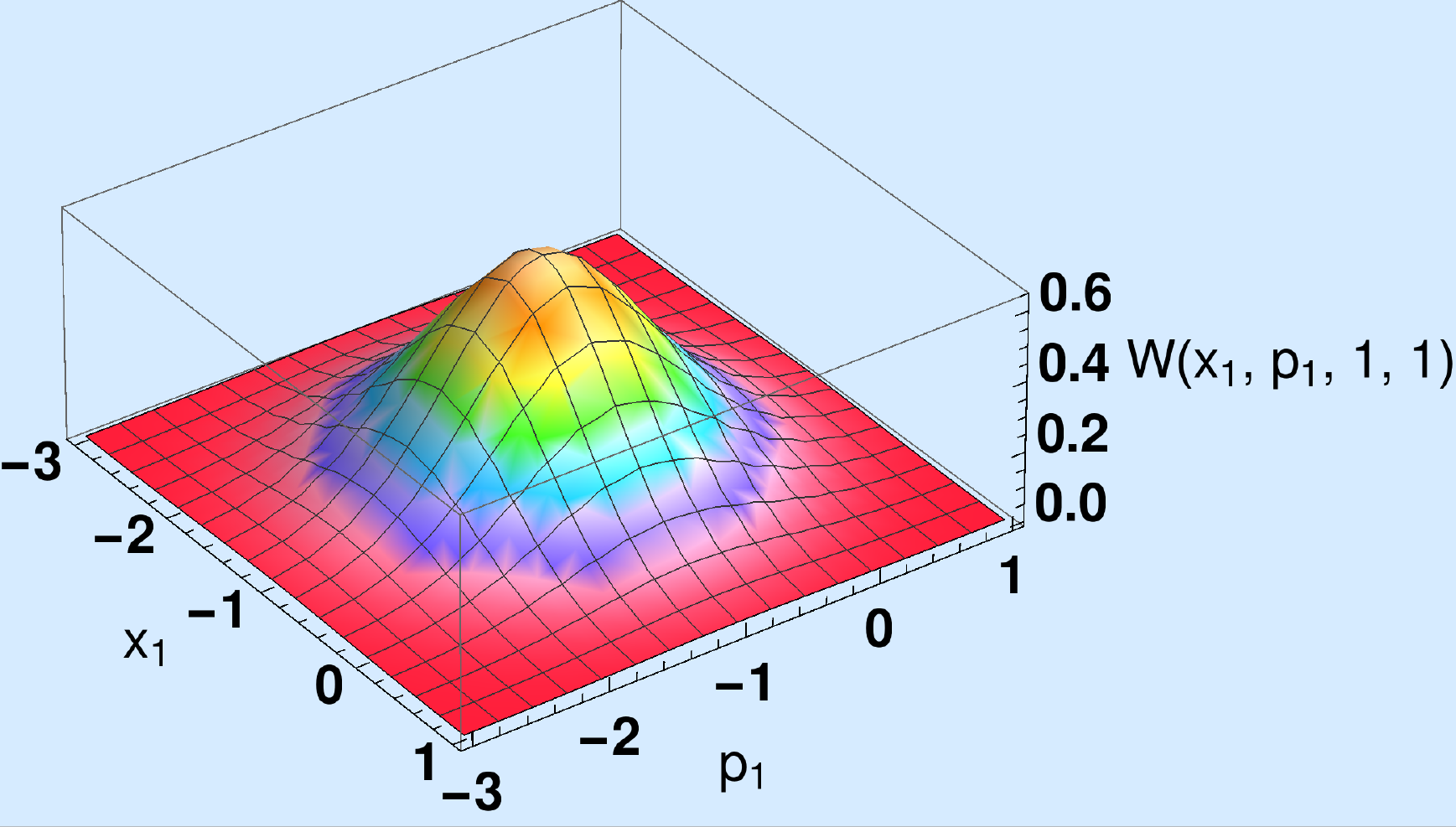}\\
{\footnotesize projection on $(x_2,p_2)=(1,1)$ plane.}
\end{minipage}
\begin{minipage}[t]{0.5\textwidth}
\centering\includegraphics[width=\textwidth]{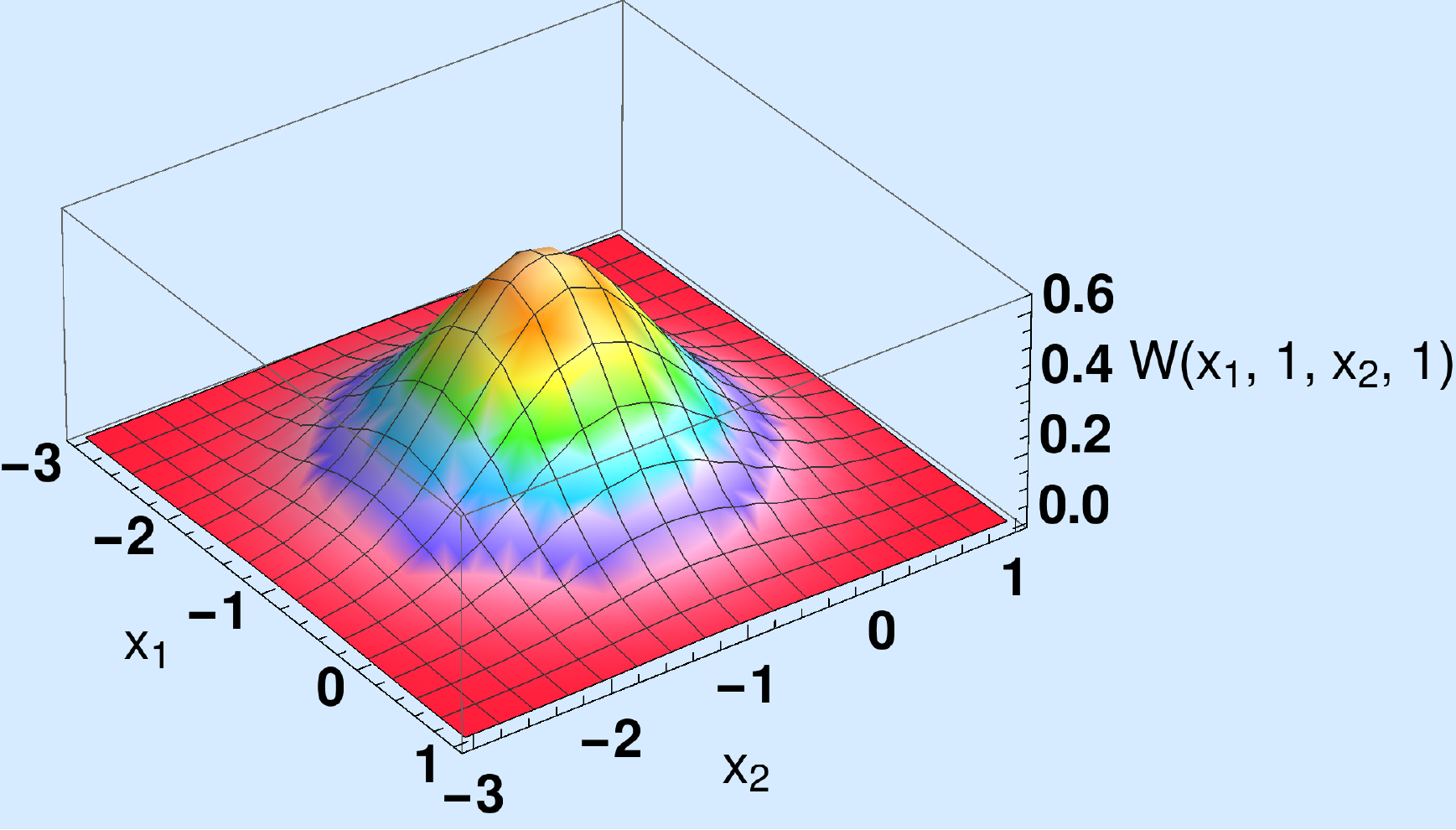}\\
{\footnotesize projection on $(p_1,p_2)=(1,1)$ plane.}
\end{minipage}\hfill
\begin{minipage}[t]{0.5\textwidth}
\centering\includegraphics[width=\textwidth]{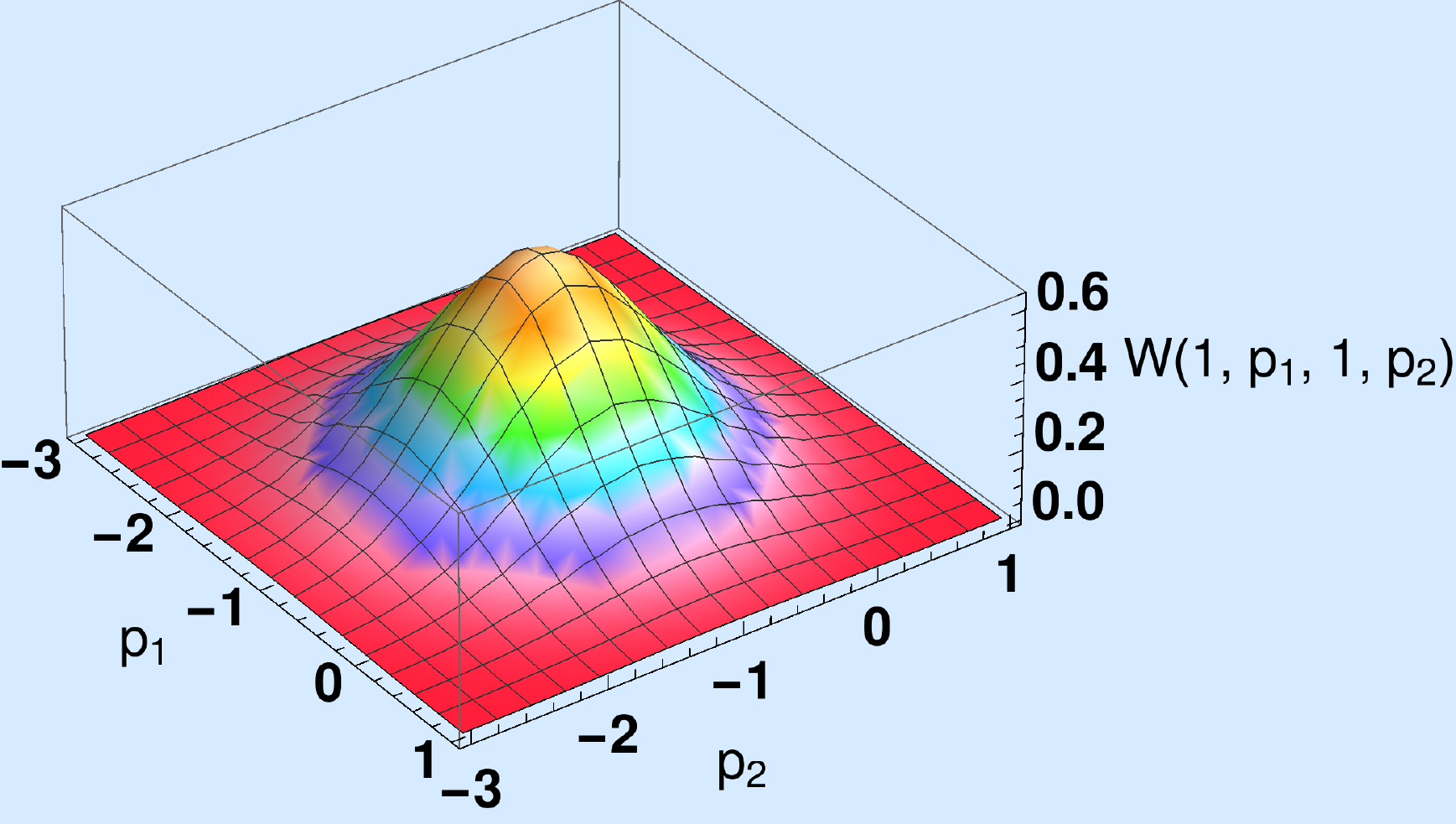}\\
{\footnotesize projection on $(x_1,x_2)=(1,1)$ plane.}
\end{minipage}\hfill
\caption{Wigner distribution $W(1,1,x_2,p_2)$, $W(x_1,p_1,1,1)$, $W(x_1,1,x_2,1)$ and $W(1,p_1,1,p_2)$ shows no entanglement effect. }\label{sampleFig1}
\end{figure}
\begin{figure}
 \begin{minipage}[t]{0.5\textwidth}
  \centering\includegraphics[width=\textwidth]{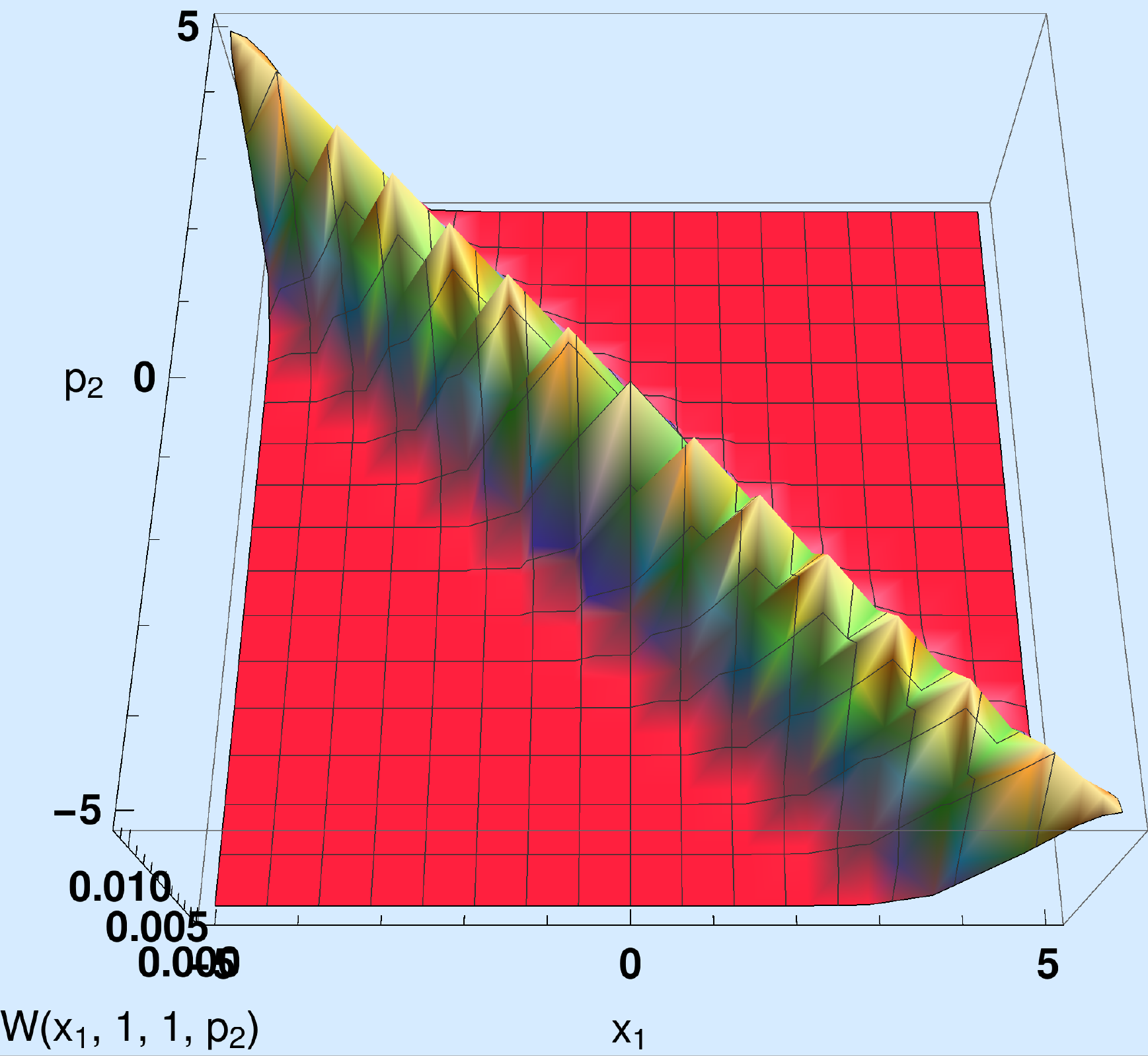}\\
  {\footnotesize Projection on $(p_1,x_2)=(1,1)$ plane}
 \end{minipage}\hfill
 \begin{minipage}[t]{0.5\textwidth}
  \centering\includegraphics[width=\textwidth]{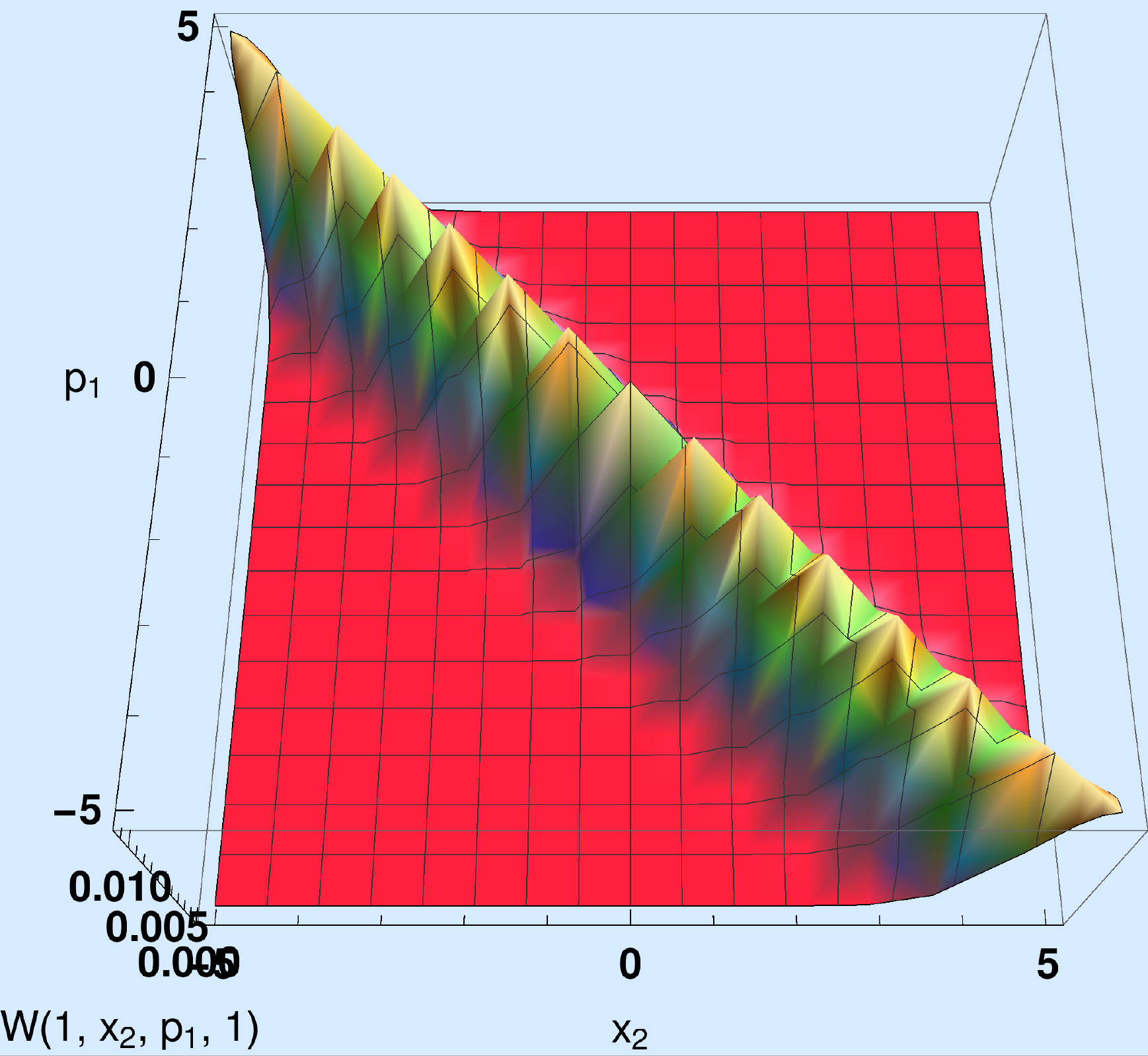}\\
  {\footnotesize Projection on $(x_1,p_2)=(1,1)$ plane}
 \end{minipage}\hfill
\caption{Wigner distribution $W(x_1,1,1,p_2)$ and $W(1,x_2,p_1,1)$ are exhibiting the entanglement effects. They are not appearing as a separated Gaussian distribution.}\label{sampleFig2}
\end{figure}
For the illustration purpose, let us choose 
\begin{eqnarray}
 \langle \hat{x}_1^2 \rangle = \langle \hat{x}_2^2 \rangle =  \langle \hat{p}_1^2 \rangle =  \langle \hat{p}_2^2 \rangle =\frac{1}{2}, \nonumber \\
 \langle \hat{x}_1\hat{p}_2 \rangle = \langle \hat{p}_1\hat{x}_2 \rangle =-\frac{1}{2}.
\end{eqnarray}
Since our $W$ is on four-dimensional phase space, we have to project it onto two-dimensional space to visualize it graphically. 
The FIG.~\ref{sampleFig1} represents the projection of WQD onto $(x_1,p_1)=(1,1)$, $(x_2,p_2)=(1,1)$, $(p_1,p_2)=(1,1)$ and $(x_1,x_2)=(1,1)$ planes respectively.  They exhibit no entanglement effect. Being projected onto the two-dimensional phase-space, they appear to be an individual Gaussian distribution. 
However, from FIG.~\ref{sampleFig2}, one can envisage the entanglement of the co-ordinates $(x_1,p_2)$ and $(x_2,p_1)$. In particular, $x_1$ is entangled with $p_2$ and $x_2$ is entangled with $p_1$. 
This confirms the indications we had to get from the correlations in the previous section. In particular, the figures are consistent with the correlations matrix (variance matrix) of the last section. 
\section{Experimental connections}
\subsection{Szilard engine cycle}
One of the possibilities of results of the present study is to use it  for the Szilard engine cycle \cite{szilard,szilard1,szilard2,szilard3,szilard4,szilard5} for a bipartite system.  Gaussian measurement performed by the second party (corresponding to $x_2$) corresponds to the variance matrix \cite{measurementgaussian} 
\begin{equation}
 \hat{\gamma}^{\pi_2}= \hat{R}(\theta_{\pi_2})\mbox{diag}(\mu/2,\mu^{-1}/2)\hat{R}^T(\theta_{\pi_2}),\;\; \mu\in [0,\infty).
\end{equation}
Where the rotation operator is given through the Pauli matrix by $\hat{R}(\theta_{\pi_2})=\cos\theta_{\pi_2}\hat{\mathbb{I}} -i\sin\theta_{\pi_2}\hat{\sigma}_y$. The measurement parameter $\mu=0$ and $\mu=1$ corresponds to homodyne  and heterodyne  measurement respectively. The conditional state covariance matrix is given by
\begin{equation}
 \hat{V}_{11}^{\pi_2}= \hat{V}_{11} -\hat{V}_{12}(\hat{V}_{22}+ \hat{\gamma}^{\pi_2})^{-1}\hat{V}_{12}^T .
\end{equation}
Then the extractable work ($W_{se}=K_B T [S(V_{11})-S(V_{11}^{\pi_2})]$) due to the measurement backreaction is given by
\begin{equation}
 W_{se}=\frac{1}{2}k_BT\ln\left(\frac{det(\hat{V}_{11})}{det(\hat{V}_{11}^{\pi_2})}\right),
\end{equation}
where we have used the R\'{e}nyi entropy of order $2$, $S_2(\rho)=-\ln (Tr(\rho^2))$, which in case of Gaussian state becomes $S_2(V_{12})=\frac{1}{2}\ln(det (V_{12}))$.\\
In the concerned case of this present paper, for a heterodyne measurement ($\mu=1$), the extractable  work is reduecd to
\begin{equation}
 W_{se}=-\frac{1}{2}k_B T \ln\left[ \left(1+\frac{\Lambda_{12}^2}{(d+2\Lambda_{11})}\right) \left(1+\frac{\Lambda_{12}^2}{d(1+2\Lambda_{22})}\right) \right] \neq 0.
\end{equation}
In particular, the non-commutativity of the space opens up a possibility of extractable work protocol.
\subsection{Determination of the signature of noncommutative of space: an Opto-mechanical scheme }
The optomechanical scheme is realized through an interaction between an optical pulse and a mechanical oscillator \cite{expt1,expt2}. The setup  consists of a half-silvered mirror, two Fabry-Perot cavities,
a laser source, and a detector (for instance see \cite{expt1,expt2,expt3}), on which a sequence of
four radiation pressure interactions form an optical cavity (resonator) yielding the  interaction operator 
\begin{equation}
 \hat{\zeta}_{lm}=\hat{U}_{lm,p}\hat{U}_{lm,x}^\dagger\hat{U}_{lm,p}^\dagger \hat{U}_{lm,x},\; \mbox{with}\; \hat{U}_{lm,p}=e^{i\hat{n}_l\sum_{j=1}^{2} \lambda_{jl}\hat{\tilde{P}}_j^m},\; \hat{U}_{lm,x}=e^{i\hat{n}_l\sum_{j=1}^{2} \lambda_{jl}\hat{\tilde{X}}_j^m},
\end{equation}
where the index $m$ and $l$ stand for the identification of the mechanical quantities and the optical degrees of freedom, respectively, and $\lambda_l=(\lambda_{1l},\lambda_{2l})$ is the optical field interaction length. $\hat{\zeta}_{lm}$ displaces the mechanical state around complete loop in phase space, thus
creating an additional phase, which is measurable in a tabletop experiment, in principle. In particular, the mean of the optical field operator
\begin{equation}
 \langle \hat{a}\rangle = \langle \alpha\vert  \hat{\zeta}_{lm}^\dagger \hat{a} \hat{\zeta}_{lm}\vert \alpha\rangle =  \langle \hat{a}\rangle_{QM} e^{-i\Theta}
\end{equation}
provides the additional phase ($\Theta$) on the light arising due to the presence of the noncommutative oscillator. Whereas, the mean of the annihilation operator $\hat{a}$ of the optical field $\vert \alpha\rangle$ (the input optical coherent state with mean photon number $n_p$) using the standard quantum mechanics (with commutative space) reads
\begin{equation}
 \langle \hat{a}\rangle_{QM} = \alpha e^{-i\vert \lambda_l\vert^2 -n_p(1-e^{-2i\vert \lambda_l\vert^2})}.
\end{equation}
The additional phase factor $\Theta$ depends on the product of the noncommutative parameters ($\theta\eta$). That means the nonzero $\Theta$ implies both position-position and momentum-momentum noncommutativity.
\section{Conclusions}
In this paper, we have studied in detail the entanglement property of two
dimensional anisotropic (a fairly general form of anisotropy) harmonic oscillator (AHO, where the entanglement is induced by both the spatial and momentum non-commutativity.
It turns out that the bipartite Gaussian state (BGS) is almost always entangled. Simon's separability criterion shows that BGS is separable for a set of specific values of the masses and frequencies of the oscillators. The explicit form of the constraint equation is determined in the present paper.
  An additional interaction term, dependent upon
the noncommutativity parameter $\theta$ and $\eta$, is generated by Bopp's shift, which is the main source for the entanglement between the coordinates and momentum degrees of freedom. \\
Wigner quasiprobability distribution (WQD) for the ground state of our system is computed. Projecting the four-dimensional phase space WQD onto two dimensions, we have graphically identified the exact degrees of freedom, which are entangled. This is also demonstrated by calculating the correlations (variances) between the observables. This opens up the possibility of direct experimental observation through interferometry. In particular, the obtained coherent state structure can be utilized in an optomechanical setup that helps us to transfer the information of a noncommutative anisotropic oscillator to the high-intensity optical pulse in terms of a sequence of opto-mechanical
interaction inside an optical resonator \cite{expt1,expt2,expt3,expt4,expt5}. Consequently, the optical phase shift is easily measurable with very high accuracy through an interferometric system as stated in \cite{expt1}. This makes the whole procedure much easier to collect the pieces of information of noncommutative structures through optical systems already available to us. In particular, by comparing the phase difference of the input signal and output signal after interaction of the input signal with the mechanical anisotropic oscillator inside the cavity, and comparing the phase difference with the prediction of usual quantum mechanics, we can determine the signature of quantum gravity  (if any). Thus, it may bypass all the difficulties of probing high energy scales through scattering experiments.  However, we would like to mention that, within this type of experimental set-up only the product $\theta\eta$ are determined, not the individual $\theta$ and $\eta$. Moreover,  the typical size ( distance between mirrors) of the resonant cavity for a Fabry-Perot Interferometer should be an order of a few kilometers to have any detectable effect in this scenario \cite{expt1}.\\
Moreover, the amount of extractable work for a heterodyne measurement through Szilard engine cycle for our bipartite system is outlined in this paper. This shows the applicability of our results in thermodyanmic process, i.e, in engineering applications.\\
In terms of constructor-theoretic principles \cite{marletto1,bose1}, if we observe the entanglement effects in the measurement of the properties of two quantum masses that interact with each other through gravity only, then we can conclude that the mediator (gravity) got to have some quantum features \cite{marletto1,bose1}. It doesn’t matter in what way gravity is quantum - whether it’s loop quantum gravity or string theory or something else - but it has to be a quantum theory. We have seen in the present paper that the entanglement in co-ordinate degrees of freedom is solely depends on the noncommutative parameters. Thus any experimental signature of entanglement for the present scenario will indicate the signature of noncommutativity of space.   

 \section{Acknowledgement}
 I am grateful to the anonymous referees for their valuable suggestions.\\
 This research was supported in part by the International Centre for Theoretical Sciences (ICTS) for the online program - Non-Hermitian Physics (code: $ICTS/nhp2021/3$).
 \section{Conflict of interest}
The author declares that there is  no conflict of interest for the present article.
\section{Data availability statement}
Data sharing is not applicable to this article as no new data were created or analyzed in this study.

\begin{appendices}
\section{Diagonalization of $\hat{\Omega}$}
$\hat{\Omega}$ is a normal matrix.  So, $\hat{\Omega}$  is diagonalizable through a   similarity transformation 
\begin{equation}\label{diagonalizationofOmega}
 \hat{\Omega}_D= \hat{Q}^{-1} \hat{\Omega}\hat{Q}.
\end{equation}
  $\hat{Q}^{-1}$ and $\hat{Q}$ are obtained by arranging the left and right eigenvectors column-wise, respectively.
Since, $\hat{\Omega}$ is not a symmetric matrix ($\hat{\Omega}^T \neq \hat{\Omega}$), the left and right eigenvectors of $\hat{\Omega}$ are not same. However,  the left and right eigenvalues are the same.  The characteristic polynomial  
($
 p(\lambda)= \det (\lambda \mathbb{\hat{I}}- \hat{\Omega}) ,
$)
of  $\hat{\Omega}$ is given by
\begin{equation}
 p(\lambda)= \lambda^4+ b\lambda^2 +c .
\end{equation}
Where

\begin{eqnarray}
 c &=& \omega_x^2 \omega_y^2 ,\\
 b &=& \alpha_0 \omega_x^2 + \frac{1}{\alpha_0}\omega_y^2 + 4\nu_1\nu_2 (\sqrt{\alpha_0}+\frac{1}{\sqrt{\alpha_0}})^2 .
\end{eqnarray}
With
\begin{eqnarray}
 \omega_x^2 &=& 4\nu_1\nu_2 (\frac{\mu_1\omega_1^2}{4 \mu_2\nu_2^2} -1) ,\\
 \omega_y^2 &=& 4\nu_1\nu_2 (\frac{\mu_2\omega_2^2}{4 \mu_1\nu_1^2} -1), \\
 \alpha_0 &=& \frac{\mu_2\nu_2}{\mu_1\nu_1}.
\end{eqnarray}
using the explicit forms of $\mu_i,\nu_i,\omega_i,\; i=1,2$, one can show that 
$
 c \ge 0,\; b\ge 0
$ as follows.
\subsection{Proof of $b\ge 0$, $c\ge 0$}
We have
\begin{eqnarray}
 b &=& \omega_1^2 + \omega_2^2 + 6\nu_1\nu_2 ,\\
 c &=& \omega_1^2  \omega_2^2 + 16 \nu_1^2 \nu_2^2 -4\nu_1^2 \omega_1^2 \mu_1/\mu_2 - 4\nu_2^2 \omega_2^2 \mu_2/\mu_1 .
\end{eqnarray}
Clearly, $b>0$ for nonzero positive values of $\omega_1, \omega_2, \nu_1, \nu_2$. \\
First we observe that, $c$ can be factorized as
\begin{equation}
 c= 16\nu_1^2\nu_2^2 \left( \frac{\mu_1\omega_1^2}{4\mu_2\nu_2^2}-1 \right) \left( \frac{\mu_2\omega_2^2}{4\mu_1\nu_1^2}-1 \right).
\end{equation}
If we use the explicit forms of $\omega_1, \omega_2, \nu_1, \nu_2$, we have
\begin{eqnarray}
 \frac{\mu_1\omega_1^2}{4\mu_2\nu_2^2} &=& \frac{4m_2^2\hbar^2}{(\eta+ m_1m_2\theta \tilde{\omega}_1^2)^2} \left(m_1\tilde{\omega}_1^2+ \frac{\eta^2}{4\hbar^2 m_2}\right) \left( \frac{1}{m_2} + \frac{1}{4\hbar^2}m_1\tilde{\omega}_1^2 \theta^2 \right) \nonumber \\
 &=& \left(1+ \frac{4\hbar^2}{\eta^2}m_1m_2 \tilde{\omega}_1^2 \right)  \left(1+ \frac{\theta^2}{4\hbar^2}m_1m_2 \tilde{\omega}_1^2 \right) \left(1+ \frac{\theta}{\eta}m_1m_2 \tilde{\omega}_1^2 \right)^{-2} \nonumber \\
 &=& \frac{1}{\left(1+ \frac{\theta}{\eta}m_1m_2 \tilde{\omega}_1^2 \right)^{2}}\left( \left(1+ \frac{\theta}{\eta}m_1m_2 \tilde{\omega}_1^2 \right)^{2} + \frac{2\theta}{\eta}m_1m_2\tilde{\omega}_1^2 (\frac{2\hbar^2}{\eta\theta}+ \frac{\eta\theta}{8\hbar^2}-1)\right) \nonumber \\
 &=& 1+ \frac{2\theta m_1m_2 \tilde{\omega}_1^2}{\eta \left(1+ \frac{\theta}{\eta}m_1m_2 \tilde{\omega}_1^2 \right)^{2}} \left( \frac{\sqrt{2}\hbar}{\sqrt{\eta\theta}} - \frac{\sqrt{\eta\theta}}{2\sqrt{2}\hbar} \right)^2  
  \ge 1 .
\end{eqnarray}
Similarly, we see that
\begin{equation}
 \frac{\mu_2\omega_2^2}{4\mu_1\nu_1^2} = 1+ \frac{2\theta m_1m_2 \tilde{\omega}_2^2}{\eta \left(1+ \frac{\theta}{\eta}m_1m_2 \tilde{\omega}_2^2 \right)^{2}} \left( \frac{\sqrt{2}\hbar}{\sqrt{\eta\theta}} - \frac{\sqrt{\eta\theta}}{2\sqrt{2}\hbar} \right)^2  \ge 1.
\end{equation}
Therefore, we have
\begin{equation}
 c\ge 0.
\end{equation}

Moreover, the discriminant of the characteristic polynomial 
\begin{eqnarray}
 \Delta &=& b^2 -4c \nonumber \\
 &=& \left(\alpha_0 \omega_x^2 -\frac{1}{\alpha_0}\omega_y^2\right)^2 + 8\nu_1\nu_2 (1+ \alpha_0)^2 (\omega_x^2 +\omega_y^2) + \frac{4\nu_1^2 \nu_2^2}{\alpha_0^2} (1+\alpha_0)^4 
\end{eqnarray}
is positive definite. It can be easily verified that
\begin{equation}\label{lambdasquare}
 \lambda^2 = \frac{1}{2}(-b\pm \sqrt{\Delta}) \le 0.
\end{equation}
\subsection{Eigenvalues and eigenvectors of $\hat{\Omega}$}
 From ~\eqref{lambdasquare}, we can see that $\hat{\Omega}$ has four distinct imaginary eigen-values
\begin{eqnarray}
 \lambda=\left\{ -i\lambda_1,i\lambda_1, -i\lambda_2,i\lambda_2 \right\}.
\end{eqnarray}
Where
\begin{equation}
\lambda_1 = \frac{1}{\sqrt{2}}\sqrt{b+\sqrt{\Delta}} ,\;
\lambda_2 = \frac{1}{\sqrt{2}}\sqrt{b-\sqrt{\Delta}} .
\end{equation}
Let us consider the eigenvalue equations of $\hat{\Omega}$ as follows.
\begin{eqnarray}
 u_j \hat{\Omega} &=& -i\lambda_j u_j,\; j=1,2.\\
 u_j^* \hat{\Omega} &=& i\lambda_j u_j^*\;, j=1,2.\\
 \hat{\Omega}v_j &=& -i\lambda_j v_j,\; j=1,2. \\
 \hat{\Omega}v_j^* &=& i\lambda_j v_j^*,\; j=1,2.
\end{eqnarray}
Then the following identities hold quite generally.
\begin{equation}\label{orthogonalityofuv}
 u_i^* v_j = u_i v_j^*,\;
 u_i v_j = u_i^* v_j^* =\delta_{ij},\; \forall i,j=1,2.
\end{equation}
~\eqref{orthogonalityofuv} suggests the following  conditions  between the left and right eigen-vectors.
\begin{equation}\label{leftrightconnection}
 v_i= -\Sigma_y u_i^\dagger, \; (i=1,2).
\end{equation}

In our case, the explicit form of the left eigenvectors $u_i,\;(i=1,2)$ are given by
\begin{eqnarray}
 u_i = \frac{1}{k_i} \left(\begin{array}{c}
-i\lambda_i \mu_1 \mu_2 (\lambda_i^2 -\omega_2^2 -4\nu_1\nu_2) \\
                            \mu_2(\lambda_i^2 -\omega_2^2)+4\mu_1\nu_1^2 \\
                            2\nu_1 \mu_1\mu_2 (\lambda_i^2 -4\nu_1\nu_2)+ 2\nu_2 \mu_2^2 \omega_2^2 \\
2i\lambda_i(\mu_1\nu_1 +\mu_2\nu_2)
                           \end{array}
\right)^T.
\end{eqnarray}
The right eigenvectors can be constructed from ~\eqref{leftrightconnection} in a straighforward manner. 
\subsection{Similarity transformation}
Now the similarity transformation $\hat{Q}$ (as well as $\hat{Q}^{-1}$) can be constructed by arranging the  eigen-vectors columnwise. In particular, 
\begin{eqnarray}
 \hat{Q}&=& (v_1,v_1^*,v_2,v_2^*), \\
 \hat{Q}^{-1} &=& (u_1^T,(u_1^*)^T,u_2^T,(u_2^*)^T).
\end{eqnarray}
One can  verify that 
\begin{equation}\label{Qdagger}
 \hat{Q}^\dagger = -\hat{\Sigma}_z \hat{Q}^{-1} \hat{\Sigma}_y,
\end{equation}
with
\begin{equation}
 \hat{\Sigma}_z = \mbox{diag}(\sigma_z,\sigma_z).
\end{equation}

Now, using ~\eqref{diagonalizationofOmega}, we have the diagonalization representation $\hat{\Omega}_D$ of $\hat{\Omega}$ as
\begin{equation}
 \hat{\Omega}_D= \mbox{diag}(-i\lambda_1,i\lambda_1,-i\lambda_2,i\lambda_2).
\end{equation}
The similarity transformation suggests a new co-ordinate vector $\hat{\zeta}$, which  is connected with the original co-ordinate vector $X$ of ~\eqref{Xdefn} through
\begin{equation}\label{zetadefn}
 \hat{\zeta}= \hat{Q}^{-1}X.
\end{equation}
Then, using ~\eqref{Qdagger}, we can write
\begin{eqnarray}
 X^\dagger &=&  \hat{\zeta}^\dagger (-\hat{\Sigma}_z \hat{Q}^{-1} \hat{\Sigma}_y) .
\end{eqnarray}
\subsection{Equivalent dynamics in diagonal representation}
The time-dependent Schr\"{o}dinger equation 
\begin{equation}
 \hat{H}\psi =i\hbar \frac{\partial \psi}{\partial t},
\end{equation}
is transformed with the similarity transformation as follows.
\begin{eqnarray}
 \hat{Q}^{-1} \frac{1}{2} X^\dagger \hat{\mathcal{H}} X \psi \hat{Q} &=& i\hbar \hat{Q}^{-1} \frac{\partial \psi}{\partial t} \hat{Q}.\\
 \therefore \frac{1}{2} \hat{\zeta}^\dagger \Sigma \hat{\zeta} \tilde{\psi}_Q &=& i\hbar \frac{\partial \tilde{\psi}_Q}{\partial t} .
\end{eqnarray}
Where
\begin{eqnarray}
 \hat{\Sigma}
  = \mbox{diag}(\lambda_1,\lambda_1,\lambda_2,\lambda_2), \label{Hdiag}\\
  \tilde{\psi}_Q =  \hat{Q}^{-1} \psi  \hat{Q}.
\end{eqnarray}
That means, the new diagonal Hamiltonian 
\begin{equation}
 \hat{H}_D= \frac{1}{2}\zeta^\dagger \hat{\Sigma}\zeta
\end{equation}
obeys the same dynamics as that of $\hat{H}$, with the new co-ordinate vector $\zeta$.\\
If we compute the commutators of the components of $\hat{\zeta}$, then we can identify that they forms a set of independent annihilation and creation operators. In particular,
\begin{equation}
 \left[\hat{\zeta}_k, \hat{\zeta}_l^\dagger\right]=\delta_{kl};\; k,l=1,2.
\end{equation}
Where
\begin{equation}
 \hat{\zeta} =(\hat{\zeta}_1, \hat{\zeta}_1^\dagger, \hat{\zeta}_2, \hat{\zeta}_2^\dagger)^T.
\end{equation}
\end{appendices}

\end{document}